\documentclass[mathleft]{an}
\usepackage{graphicx}
\usepackage{aas_macros}
\usepackage{times}
\usepackage{epsf}
\usepackage{epsfig}
\usepackage{natbib}
\bibpunct{(}{)}{;}{a}{}{,}
\begin{document}

\Pagespan{1057}{1060}
\Yearpublication{2012}%
\Yearsubmission{2012}%
\Month{11}%
\Volume{333}%
\Issue{10}%
\DOI{10.1002/asna.201211804}%

\newcommand{\kepler}{\textit{Kepler}}

\title{Kepler Fourier Concepts: the performance of the Kepler data pipeline.}

\author{Simon J. Murphy\inst{1}\fnmsep\thanks{Corresponding author: {smurphy6@uclan.ac.uk}\newline}}
\titlerunning{Kepler Fourier Concepts}
\authorrunning{S.J. Murphy}
\institute{
\inst{1}Jeremiah Horrocks Institute, University of Central Lancashire, Preston PR1 2HE
}

\received{2012 Oct 18}
\accepted{2012 Oct 22}
\publonline{2012 Dec 3}


\abstract{Given the extreme precision attainable with the Kepler Space Telescope, the mitigation of instrumental artefacts is very important. In an earlier paper \citep{murphy2012}, the characteristics of \kepler\ data were discussed in light of their effect on asteroseismology. We continue this discussion now that data processed with the new PDC-MAP pipeline are publicly available; users should use the latest data reductions available at the Mikulski Archive for Space Telescopes (MAST), not just for PDC, but also for improvements in the attached meta-data. We discuss the injection of noise in the frequency range 0-24\,d$^{-1}$ (up to $\sim$277\,$\mu$Hz) by the PDC-LS pipeline into $\sim$15\,per\,cent of light-curves.}

\maketitle

\section{Introduction}
The exquisite precision and long time-base of \kepler\ data make for excellent demonstrations of some fundamental concepts particular to Fourier transforms. Many of these were presented in \citet[][hereafter Paper\,I]{murphy2012}, which addressed characteristics of data of different sampling rates, namely of the \kepler\ short- and long-cadence data, with 58.9-s and 29.4-min exposures, respectively.

Paper\,I demonstrated some basic Fourier concepts as applied to \kepler\ data, such as the importance of having short-cadence (SC) data for investigating $\delta$\,Sct stars because of the aliasing generated by the low long-cadence (LC) Nyquist frequency, and also how the study of flares, for instance, can benefit from the increased time-resolution (see, e.g. \citealt{balona2012b}). Furthermore, Paper\,I showed how SC data allow for a more precise determination of pulsation frequencies, amplitudes and phases, but do not offer greater frequency resolution, and how the observed pulsation amplitudes in LC data suffer an amplitude-reduction effect due to under-sampling. Also presented therein was the performance of the Pre-search Data Conditioning (PDC) pipeline, whose job it is to remove instrumental systematic signatures whilst preserving the astrophysics, compared to the Simple Aperture Photometry (SAP) data, which undergo only basic calibration. It was shown that the PDC data of Data Release 11 and earlier showed much lower noise than their SAP counterparts, but that spurious low-frequency peaks were evident in the LC PDC data in non-pulsating stars.

This article offers a continuation from Paper\,I in which we further address a newly-discovered and characterised form of noise injection in the older version of the PDC pipeline, which used least-squares algorithms to process the data, and has thus been renamed PDC-LS to alleviate confusion -- this characterisation is presented in \S\,\ref{sec:noise}. An early performance review of the newer, Maximum A Posteriori PDC pipeline (PDC-MAP) is presented in \S\,\ref{sec:perf}. We discuss the benefits of continuous \kepler\ coverage in \S\,\ref{sec:cont}. The data presented in this paper pertain to Data Release 11 (or earlier) for PDC-LS, and to Data Release 14 for PDC-MAP.

\section{Noise injection by PDC-LS}
\label{sec:noise}

At the time of writing, all \kepler\ quarters have now been reprocessed with the newer PDC-MAP pipeline and are available on MAST, but PDC-MAP does not yet treat SC data, meaning that SC data are only available in the least-squares format. For further reading on the pipelines, the papers of \citet{stumpeetal2012} and \citet{smithetal2012} are recommended, along with the Kepler Data Characteristics Handbook\footnote{available at http://keplergo.arc.nasa.gov/Documentation.shtml}. In this section we provide examples of how the PDC-LS pipeline injects noise into the data, how to detect it, and the on-going efforts to handle the problem.

\begin{figure}
\begin{center}
\includegraphics[width=0.48\textwidth]{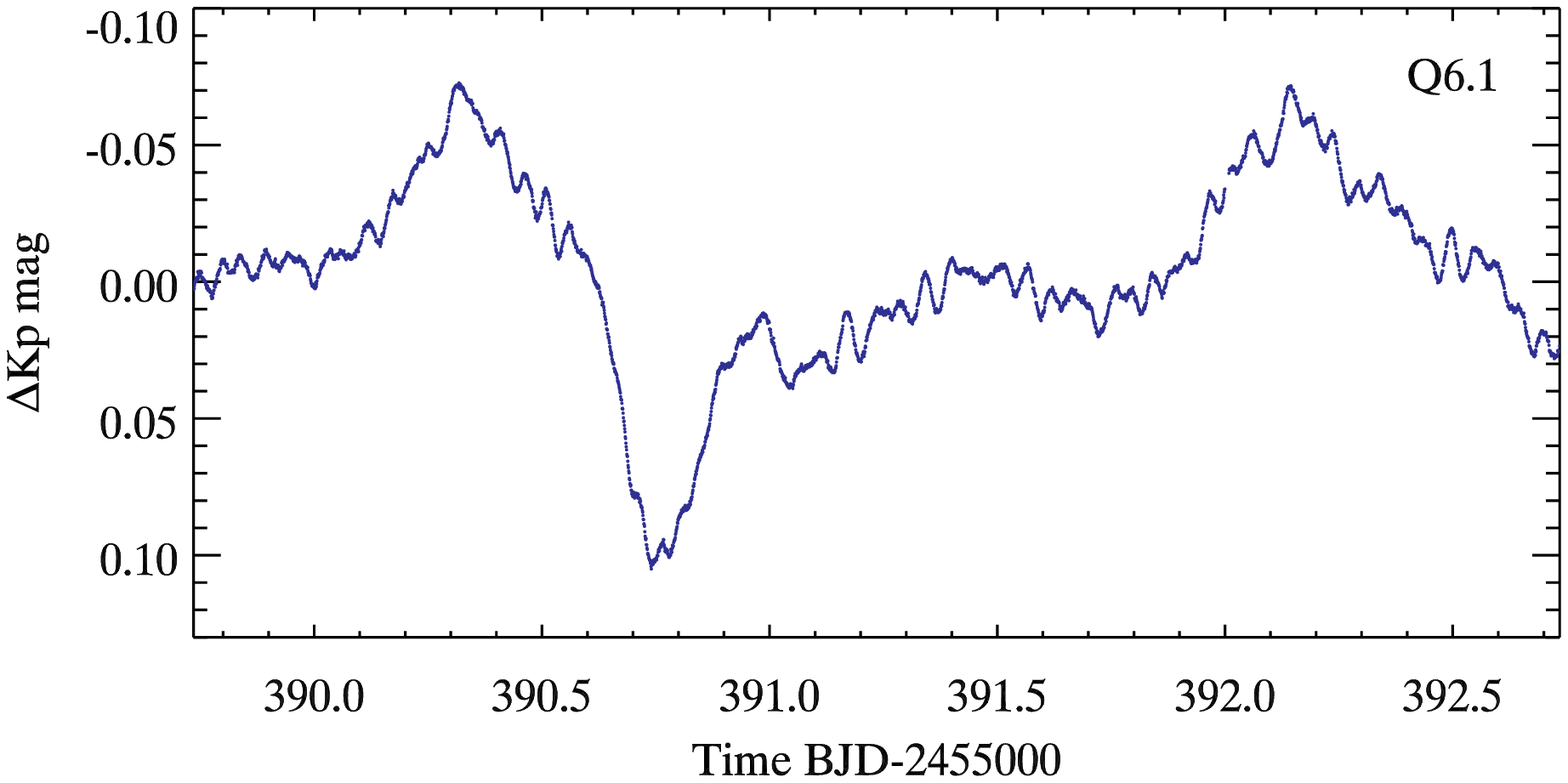}
\includegraphics[width=0.48\textwidth]{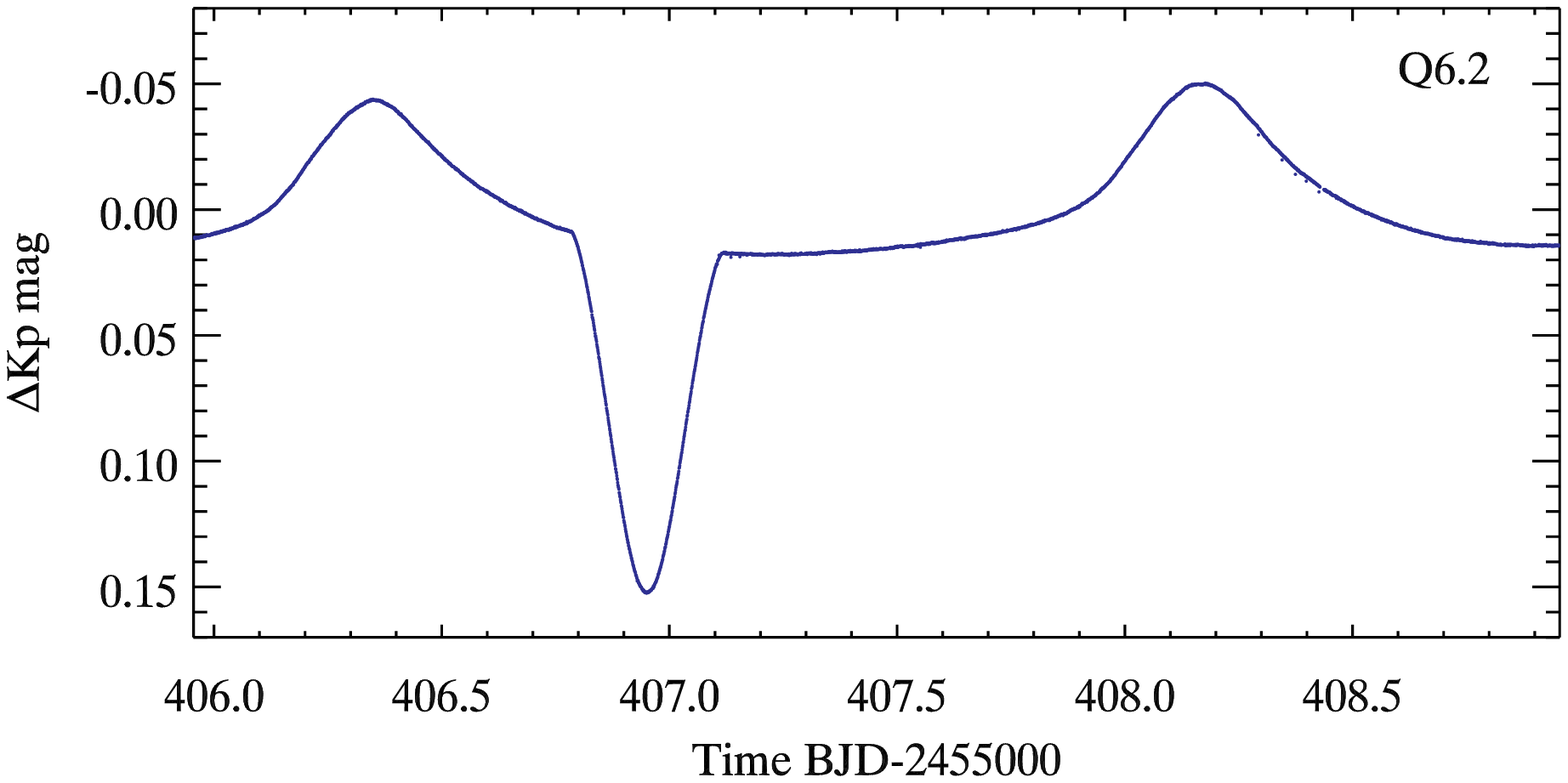}
\caption{Three-day segments of the light curves for the detached eclipsing binary system KIC\,11285625 in SC PDC flux. In Q6.1 the pipeline is injecting noise, distorting the light-curve, whereas the Q6.2 light-curve looks comparatively very clean. The corresponding Fourier transforms are shown in Fig.\,\ref{fig:FT}. It is well-known that PDC-LS does not treat binaries well, but the noise injection is not limited to eclipsing binary systems.}
\label{fig:distortion}
\end{center}
\end{figure}

\begin{figure}
\begin{center}
\includegraphics[width=0.48\textwidth]{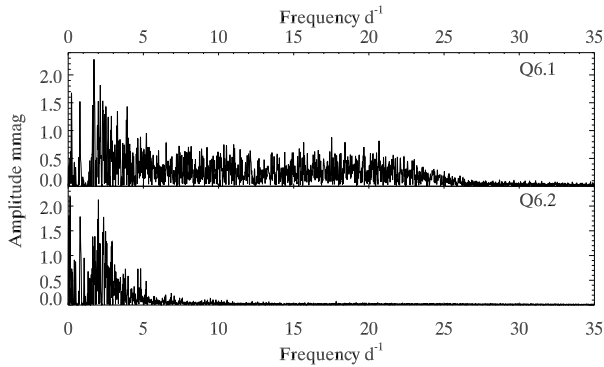}
\caption{The prewhitened Fourier transforms for Q6.1 and Q6.2 corresponding to the light curves in Fig.\,\ref{fig:distortion}, and typical for stars in which the noise injection is seen. Pulsations and many harmonics of the orbit have been prewhitened; the appearance of the Fourier transform below $\sim$5\,d$^{-1}$ is strongly affected by those harmonics that remain. \textit{Upper panel:} in the affected Q6.1 data, there is an elevated `grass level' -- the amplitude of the Fourier peaks that somewhat resemble mown grass -- up to a frequency of $\sim$24\,d$^{-1}$ after which there is a sharp drop-off in power. \textit{Lower panel:} the Q6.2 data do not suffer the noise injection. The sharp power-reduction is not noticeable in LC data (not shown) because of the low Nyquist frequency of 24.4\,d$^{-1}$.}
\label{fig:FT}
\end{center}
\end{figure}

\begin{figure}
\begin{center}
\includegraphics[width=0.48\textwidth]{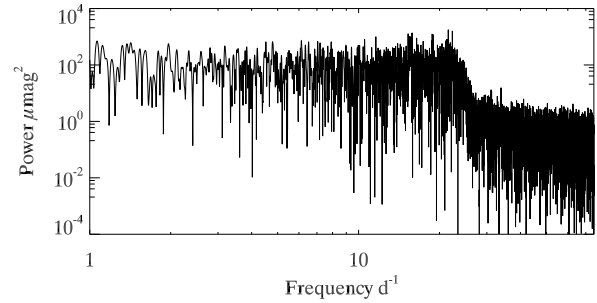}
\caption{The injected noise is easily identified by a rapid power reduction at 24\,d$^{-1}$ in the Fourier transform, plotted here in log-log space. The example given is KIC\,3429637 Q7 SC data, after pre-whitening, that is, fitting and removing the statistically significant (signal-to-noise $\geq 4$) sine curves from this $\delta$\,Sct star.}
\label{fig:log}
\end{center}
\end{figure}

We demonstrate the way in which PDC noise injection can distort light curves in Fig.\,\ref{fig:distortion}, and how this affects the Fourier transform in Fig.\,\ref{fig:FT}. One of the characteristic features of the injected noise is the drop-off in power around 24\,d$^{-1}$. Specifically, in a logarithmic plot of the power spectrum an extremely rapid reduction in power is seen (Fig.\,\ref{fig:log}), and this was used to identify stars affected by the noise injection. Indeed, this power-reduction was observed in 15\,per\,cent of the stars analysed -- those with SC data available and Kepler Input Catalogue (KIC) temperatures between 5500 and 9500\,K. The power-reduction is hard to detect in LC due to the low Nyquist frequency, but we stress that the LC data do show the injected noise.

\begin{figure}
\begin{center}
\includegraphics[width=0.48\textwidth]{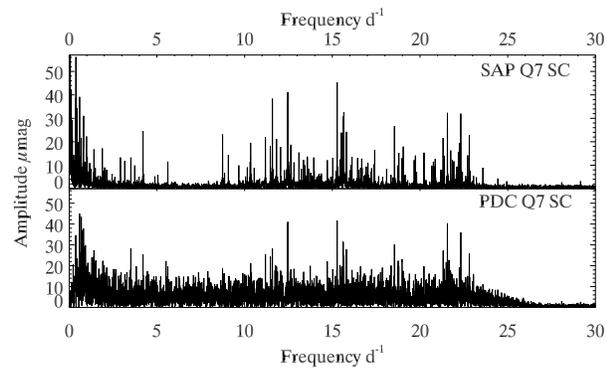}
\caption{Linear Fourier transforms for KIC\,3429637, showing the extent of the injected noise: the grass level is $\sim$15\,$\mu$mag in the PDC-LS data, compared to just 2\,$\mu$mag in the SAP data. The decrease in power at $\sim$24\,d$^{-1}$ is clear. The effect is present in Q7 in both SC and LC PDC-LS data for this star, but not present at all in Q8.}
\label{fig:3429637}
\end{center}
\end{figure}

Fig.\,\ref{fig:3429637} shows how for the star KIC\,3429637 the PDC-LS pipeline injects noise into the SC Q7 data. The noise is restricted solely to PDC-LS data and affects Q7 but not Q8 for this star. The grass level (defined in Fig.\,\ref{fig:FT} caption) is higher in PDC-LS Q7 LC than in SAP Q7 LC (after pre-whitening), so the effect is present in LC too, even if the low Nyquist frequency hides the power drop.

\begin{figure}
\begin{center}
\includegraphics[width=0.48\textwidth]{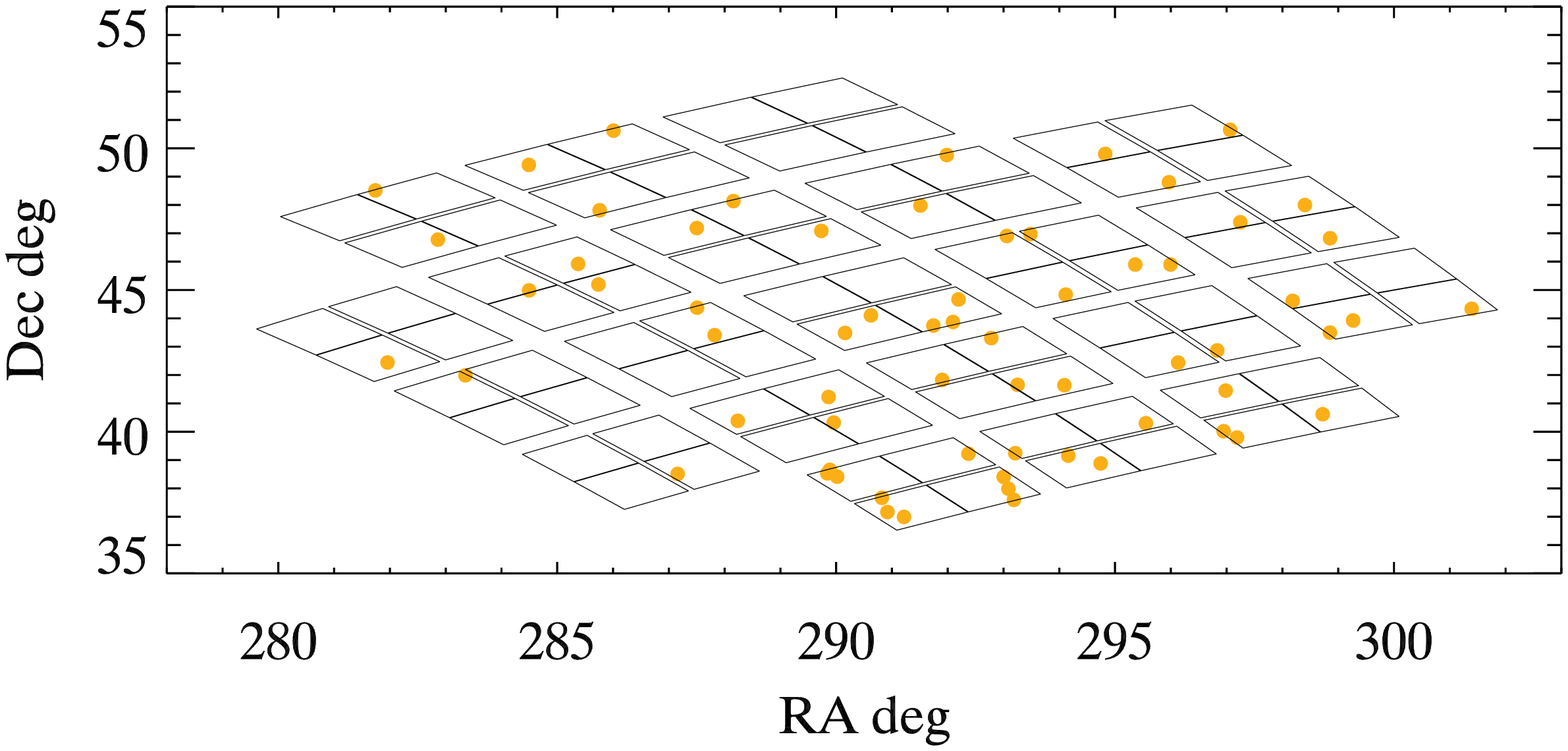}
\caption{The spatial distribution of the affected stars corresponds to the spatial density of all stars across \textit{Kepler}'s CCDs.}
\label{fig:kep_field}
\end{center}
\end{figure}

It is not known why some stars are affected while others are not, nor why, for the same star, only some quarters are affected and that these vary for different affected stars. No correlation in fraction of stars affected was seen with $T_{\rm eff}$, nor was any correlation seen with sky/CCD position (Fig.\,\ref{fig:kep_field}).

Sometimes the pipeline fails to fit a light curve and passes the light-curve through without modification. In the lower panel of Fig.\,\ref{fig:FT}, it can be seen that although noise is not injected, low-frequency peaks, which are often caused by trends in the data, are not removed either. In this case, the light curve has perhaps emerged untouched by the pipeline.

\section{Performance of PDC-MAP}
\label{sec:perf}

\begin{figure}
\begin{center}
\includegraphics[width=0.48\textwidth]{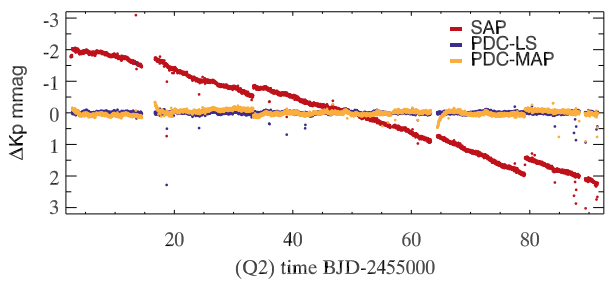}
\caption{Although PDC-MAP removes many of the long-term trends satisfactorily, it has initially struggled to mitigate thermal effects. For this reason, small exponential thermal recoveries are seen at the data downlink events in PDC-MAP, just as they are in the SAP data. The example shown is KIC\,7450391 -- a non-pulsating late A-type star, for \kepler\ Quarter 2, noting that Q2 is the most extreme example and that in other Quarters PDC-MAP does much better than this.}
\label{fig:map_lcs}
\end{center}
\end{figure}

At the time of writing, PDC-MAP has only recently become available for public data. In comparing PDC-MAP with the older PDC-LS and the SAP data (Fig.\,\ref{fig:map_lcs}), it is clear that PDC-MAP has strong trends (over many data points) following the gaps in the data that arise from the scheduled data downlinking events. In the older pipeline these trends were fixed quite well. Such teething problems might be expected with new software, somewhat similarly to how the Copernican model of the solar system initially struggled to out-perform the Ptolmiac model because of the substantially larger amount of time invested in improving the latter. PDC-MAP identifies systematic behaviour common to many stars to create a fit to subtract from the data, but there appears to be a great variation between the response of different pixels to the thermal effects that cause the trends, and this effect is convolved with differential velocity aberration, making the light-curves particularly hard to treat.

\begin{figure*}
\begin{center}
\includegraphics[width=0.99\textwidth]{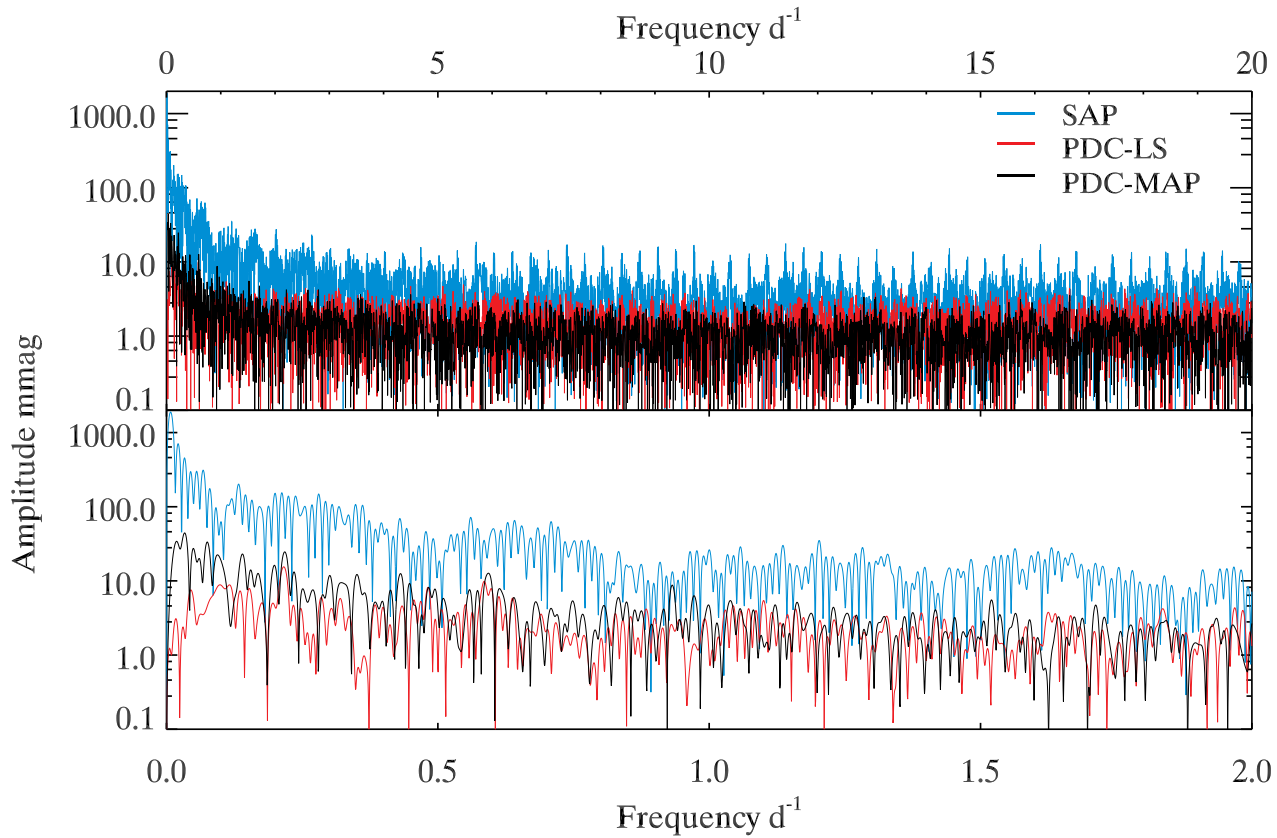}
\caption{Overlaid Fourier plot for SAP, PDC-LS and PDC-MAP for KIC\,7450391, Quarter 2, corresponding to the light-curves in Fig.\,\ref{fig:map_lcs}. \textit{Upper panel:} below $\sim$2\,d$^{-1}$ PDC-MAP is slightly noisier than PDC-LS, but has lower noise above that frequency. \textit{Lower panel:} zoom-in on the region below 2\,d$^{-1}$. SAP is substantially noisier, and the convergence in performance of PDC-MAP and PDC-LS can be seen as low as 0.5\,d$^{-1}$. PDC-MAP can thus be said to be at least as good as PDC-LS on time-scales important to astrophysics.}
\label{fig:map_FT}
\end{center}
\end{figure*}

Paper\,I investigated the difference between SAP and PDC-LS data in Fourier space, noting that the noise in the latter was reduced by factors of 10-100 over the former. Here, we extend that comparison to also include the PDC-MAP data (Fig.\,\ref{fig:map_FT}). It appears that although PDC-MAP is noisier at very low frequency due to the remaining thermal and focussing effects, at higher frequency (above $\sim$2\,d$^{-1}$) it out-performs PDC-LS, and has the added reassurance that it preserves stellar variability more satisfactorily than PDC-LS. From Quarter 13 onwards, a cubic fit will also be calculated and subtracted from the data in a pipeline version called `multi-scale' MAP. It is therein assumed that long-term trends on the scale of around one month are systematic and are thus removed.

\section{The benefits of continuous \kepler\ coverage of $\delta$\,Sct stars}
\label{sec:cont}

\kepler\ SC slots are oversubscribed, such that even though \textit{some} SC data are desirable (though not required -- to be demonstrated in a future publication) to resolve the LC Nyquist aliasing issues, it is mostly only feasible to continuously study a star in LC. Here we shall address the benefits of doing so.

The $\delta$\,Sct stars are not particularly stable pulsators. Their amplitudes and frequencies are subject to change, and the literature contains numerous examples such as 4\,CVn \citep{breger2000b}, whose pulsation amplitudes have been observed to change unpredictably for decades. With continuous, space-based observations of such high precision, the potential for observing such changes is great. \citet{murphyetal2012} have used eight Quarters of \kepler\ data to investigate the pulsational amplitude growth of the $\rho$\,Pup star KIC\,3429637. Such observations may prove indispensable in understanding mode interactions and the details of the driving mechanism and damping in operation in $\delta$\,Sct stars.

In addition, \citet{shibahashi&kurtz2012} have described a method of deriving the mass function of a binary pair using the frequency-modulation (FM) technique on Fourier data, i.e. without the need of spectroscopy. Specifically, if one has continuous LC data covering at least a whole orbit, and one of those stars pulsates (preferably at high amplitude and frequency), then by using the sidelobes present in the Fourier transform one can calculate the parameters of that orbit. So far, only three `FM stars' are known: the prototype in \citet[][KIC\,4150611]{shibahashi&kurtz2012}, a highly non-linear $\delta$\,Sct star with many combination frequencies (KIC\,11754974; Murphy et al. 2012b, in prep), and a third star currently under study by Kurtz \& Shibahashi. \citet{teltingetal2012} have found a pulsating sdB in a binary for which the sidelobes were only of borderline significance and thus direct application of the FM technique was ineffective. Under the right conditions, however, it is even possible to detect Jupiter-mass planets orbiting $\delta$\,Sct stars with this technique.

Finally, continuous coverage in eclipsing binary systems, or even transiting planets is essential to precisely determine the orbital period and even allow detections of additional bodies through Transit Timing Variations \citep{ballardetal2011}. For this reason, the news that the Kepler Space Mission is receiving funding for an extension is particularly exciting.

\section*{Acknowledgements}

SJM acknowledges the financial support of the STFC and of the ESF, and would like to thank the Kepler Science Office for their continued efforts in improving the pipeline. In particular, he would like to thank Fergal Mullally for the bilateral discussions on the pipeline.

\bibliography{sjm_2}

\begin{thebibliography}{9}
\expandafter\ifx\csname natexlab\endcsname\relax\def\natexlab#1{#1}\fi

\bibitem[{{Ballard} {et~al}\mbox{.}(2011){Ballard}, {Fabrycky}, {Fressin},
  {Charbonneau}, {Desert}, {Torres}, {Marcy}, {Burke}, {Isaacson}, {Henze},
  {Steffen}, {Ciardi}, {Howell}, {Cochran}, {Endl}, {Bryson}, {Rowe}, {Holman},
  {Lissauer}, {Jenkins}, {Still}, {Ford}, {Christiansen}, {Middour}, {Haas},
  {Li}, {Hall}, {McCauliff}, {Batalha}, {Koch}, \& {Borucki}}]{ballardetal2011}
{Ballard} S. {et~al.}, 2011, \apj, 743, 200

\bibitem[{{Balona}(2012)}]{balona2012b}
{Balona} L.~A., 2012, \mnras, 3069

\bibitem[{{Breger}(2000)}]{breger2000b}
{Breger} M., 2000, in Astronomical Society of the Pacific Conference Series,
  Vol. 203, IAU Colloq. 176: The Impact of Large-Scale Surveys on Pulsating
  Star Research, {L.~Szabados \& D.~Kurtz}, ed., pp. 421--425

\bibitem[{{Murphy}(2012)}]{murphy2012}
{Murphy} S.~J., 2012, \mnras, 422, 665

\bibitem[{{Murphy} {et~al}\mbox{.}(2012){Murphy}, {Grigahc{\`e}ne},
  {Niemczura}, {Kurtz}, \& {Uytterhoeven}}]{murphyetal2012}
{Murphy} S.~J., {Grigahc{\`e}ne} A., {Niemczura} E., {Kurtz} D.~W.,
  {Uytterhoeven} K., 2012, \mnras, 427, 1418

\bibitem[{{Shibahashi} \& {Kurtz}(2012)}]{shibahashi&kurtz2012}
{Shibahashi} H., {Kurtz} D.~W., 2012, \mnras, 422, 738

\bibitem[{{Smith} {et~al}\mbox{.}(2012){Smith}, {Stumpe}, {Van Cleve},
  {Jenkins}, {Barclay}, {Fanelli}, {Girouard}, {Kolodziejczak}, {McCauliff},
  {Morris}, \& {Twicken}}]{smithetal2012}
{Smith} J.~C. {et~al.}, 2012, \pasp, 124, 1000

\bibitem[{{Stumpe} {et~al}\mbox{.}(2012){Stumpe}, {Smith}, {Van Cleve},
  {Twicken}, {Barclay}, {Fanelli}, {Girouard}, {Jenkins}, {Kolodziejczak},
  {McCauliff}, \& {Morris}}]{stumpeetal2012}
{Stumpe} M.~C. {et~al.}, 2012, \pasp, 124, 985

\bibitem[{{Telting} {et~al}\mbox{.}(2012){Telting}, {{\O}stensen}, {Baran},
  {Bloemen}, {Reed}, {Oreiro}, {Farris}, {Ottosen}, {Aerts}, {Kawaler},
  {Heber}, {Prins}, {Green}, {Kalomeni}, {O'Toole}, {Mullally}, {Sanderfer},
  {Smith}, \& {Kjeldsen}}]{teltingetal2012}
{Telting} J.~H. {et~al.}, 2012, \aap, 544, A1

\end{thebibliography}

\end{document}